# An Exploration Study on Developing Blockchain Systems – the Practitioners' Perspective

Bakheet Aljedaani[a], Aakash Ahmad[b], Mahdi Fahmideh[c], Arif Ali Khan[d], Jun Shen[e]
[a]Computer Science Department, Aljumum University College, Umm Alqura University, Makkah, Saudi Arabia
[b]School of Computing and Communications, Lancaster University Leipzig, Germany
[c]School of Business at University of Southern Queensland, Queensland, Australia
[d]M3S Empirical Software Engineering Research Unit, University of Oulu, Oulu, Finland
[e]School of Computing and Information Technology, University of Wollongong, Wollongong, Australia
[a]bhjedaani@uqu.edu.sa, [b]a.ahmad13@lancaster.ac.uk, [c]mahdi.fahmideh@usq.edu.au, [d]arif.khan@oulu.fi, [e]jun_shen@uow.edu.au

**Abstract**

*Context:* Blockchain-based software (BBS) exploits the concepts and technologies popularized by cryptocurrencies like Bitcoin, offering decentralized transaction ledgers with immutable content for security-critical and transaction-critical systems. Recent research has explored the strategic benefits and technical limitations of BBS in various fields, including cybersecurity, healthcare, education, and financial technologies. Despite growing interest from academia and industry, there is a lack of empirical evidence, leading to an incomplete understanding of the processes, methods, and techniques necessary for systematic BBS development.
*Objectives:* Existing research lacks a consolidated view, particularly empirically-driven guidelines based on published evidence and development practices. This study aims to address the gap by consolidating empirical evidence and development practices to derive or leverage existing processes, patterns, and models for designing, implementing, and validating BBS systems.
*Method:* Tied to this knowledge gap, we conducted a two-phase research project. First, a systematic literature review of 58 studies was performed to identify a development process comprising 23 tasks for BBS systems. Second, a survey of 102 blockchain practitioners from 35 countries across six continents was conducted to validate the BBS system development process.
*Results:* Our results revealed a statistically significant difference (p-value <.001) in the importance ratings of 24 out of 26 blockchain-based software tasks by our participants. The only two tasks that were not statistically significant were incentive protocol design and granularity design. Our study also presented some of the tasks that have been emphasized by our participants within the different development phases (i.e., Analysis Phase, Design Phase, Implementation Phase, Deployment Phase, and Execution and Maintenance Phase).
*Conclusion:* Our research is among the first to advance understanding on the aspect of development process for blockchain-based systems and helps researchers and practitioners in their quests on challenges and recommendations associated with the development of BBS systems.

## 1. Introduction

Blockchain-based software (BBS) exploits the concepts and technologies popularised by cryptocurrencies such as Bitcoin – offering decentralised transaction ledgers with immutable content – to operationalise security and transaction critical systems [1]. In BBS, the concept of smart contracts enables operational transparency, user anonymity, auditability, and high scalability that interests both academic research and industrial practices on BBS-enabled platforms, technologies, and applications. A recently conducted survey [2] of more than 1400 senior executives and practitioners across the globe reveals that many IT-based organizations have found compelling use cases of blockchain systems in areas like finance, transport, healthcare, and manufacturing. Moreover, community-based development projects of BBS [3, 4] (hosted on GitHub) have seen a tremendous growth of code repositories, prototypes, and open-source initiatives that leverage blockchains for scalability and security of systems involving transaction processing, supply chain, manufacturing, medical records, and internet of things [5]. However, the unique characteristics and requirements associated with BBS systems (e.g., block mining, immutability, smart contracting, etc.) raise new challenges across their development lifecycle, entailing an extensive improvement of conventional software engineering. Engineering and development challenges correspond to ensuring the autonomy, scalability, reusability, security, auditability, governance, and other aspects that need to be addressed while modeling, developing, maintaining, and operating blockchain-based software systems [6, 7]. The development of systems leveraging blockchain platforms is a new challenging exercise for IT-based organizations and software teams involving a wide range of activities. These include, to provide a few examples, a trade-off between security and performance, the choice of consensus protocols in relation to transaction time and computing power requirements, and the cooperation of multiple institutions and stakeholders [6]. It is rational to think that developing blockchain-based systems (BBS) is, after all, essentially a type of system development.

Context – SE for BBS: Software engineering (SE) as a discipline helps researchers and practitioners to rationalise engineering knowledge and apply best practices for the design, development, validation, and deployment of existing and emerging genres of software-intensive systems effectively and efficiently [8]. In recent decades, SE focused research and development has influenced industrial processes, practices, human roles, tool support, and framework to engineer a multitude of systems ranging from cloud to mobile and internet of things. In line with this, there is a growing awareness that the end-to-end software development lifecycle requires innovative engineering approaches to bind together technical programming models (e.g., Python, Golang), platforms (e.g., Ethereum), and architectures (e.g., Hyperledger). From an engineering perspective, BBS prototypes or experimental projects may be manageable in an ad-hoc manner, however; a systematic engineering approach becomes essential if the project is extensive and/or blockchains are fundamental to support the core business processes of enterprises. Therefore, a systematic engineering and development approach can provide important activities, guidelines, checklists, design principles, and heuristics to develop BBS and mitigate, trace, and rectify operational vulnerabilities for such a class of systems. SE principles and practices can be applied to address the engineering challenges of BBS systems. In particular, 'Software engineering for Blockchain based software (SE for BBS) can help architects, engineers, and developers to tailor existing knowledge or apply innovative practices to engineer BBS systems, applications, and technologies. -intensive systems, and the need for an in-depth knowledge of cryptography, just to name a few [9]. There is a growing awareness that the end-to-end development lifecycle of BBS requires new innovative guiding engineering approaches to bind together technical programming models, platforms (e.g., Ethereum), and technologies (e.g., Bitcoin scripting languages) [3, 9, 10]. Although small BBS projects may be manageable in an ad-hoc manner, a systematic engineering approach becomes essential if a BBS project is large and the BBS is aimed to support the core business processes of an organisation.

Retrospectively, informed by the fact that the more a system gets complex, the more it opens to total breakdown [1]. This research suggests that understanding and adopting well-aligned and systematic development processes to conduct and anticipate BBS implementation is a highly critical need. Such an understanding of underlying development processes provides an overarching influence on all stages of BBS implementation lifecycle. Whilst the cogent use cases of blockchain are appealing and promise to fill a niche [3, 9], yet an integrated view underpinning the development process and its implications for BBS is non-extant in the current literature. Earlier work [6] highlighted that the current literature often suggests partial or pure technical centric solutions either applicable to only a specific stage of the BBS development or bound to one or more platforms, e.g. Ethereum, EOSIO, and Hyperledger Fabric [11]. In this work, we aim to explore what is already known about the development process for BBS, synthesize the current research and empirical data, and identify tailoring that is required to existing methods to meet blockchain development. This study examines implementing BBS from the perspective of development process models. Specifically, it seeks to address the following Research Questions (RQs):

*RQ1: How development processes (i.e., tasks/activities) are understood for blockchain systems?*
This RQ aimed to identify and categorize the various tasks and activities involved in the development process of BBS.

*RQ2: How are the existing methods (in-house or commercial) tailored to fit the development of blockchain systems?*
This RQ aimed to assess how existing software development methods, whether in-house or commercial, are adapted and tailored specifically for BBS.

*RQ3: What are the key differences between the development of conventional (non-blockchain) software systems and blockchain-based systems?*
This RQ aimed to identify and describe the distinct development processes specific to conventional software systems and BBS.

In addressing these RQs, we applied a mixed-methods research approach organized in two exploratory and confirmatory phases. In the first phase, i.e., conceptual to empirical, we drew from the existing blockchain literature a generic platform-independent process model of key phases and tasks incorporated into a typical development process of blockchain systems. The second phase, i.e., empirical to conceptual, examines the credibility of the resultant process model through conducting a survey of blockchain domain experts. We received responses from 102 randomly selected research participants with a wide range of professional backgrounds. None of the prior studies explicitly have narrowed their focus nor have empirically provided an empirical investigation on the aspect of development process for BBS. Therefore, we believe our work as the first study that explicitly unpacks the notion of development process for BBS, obtains empirical feedback from domain experts, and has significant implications for implementing BBS in real-world-scenarios. Our study provides a description of how development processes for BBS are understood, which makes it a pioneer of revelatory research in its kind in turn. Our findings help organizations and system development teams to carry out a safe and risk-aware blockchain development instead of an ad-hoc development leading to a poor and unsecured transformation to blockchain. This study offers the following primary contributions:

- Providing a generic process model including critical phases and tasks that are sequenced into implementation of systems leveraging blockchain platforms.
- Sharing quantitative and qualitative perception of blockchain domain experts about the proposed process models as well as challenges and tailoring recommendations to existing system development methods to meet blockchain implementation endeavors.
- Highlighting future research opportunities and unaddressed challenges on the aspect of development processes for BBS.

Based on the outlined contributions above, we believe our study can also have the following implications for researchers and practitioners of BBS systems:

- Enhance the understanding and practices of BBS development, offering researchers deeper insights into effective processes and challenges, and providing practitioners with practical guidelines and expert recommendations to optimize their BBS projects.
- Bridging the gap between theory and practice by providing a process model, expert insights, and future research opportunities, translating academic findings into actionable strategies. This encourages continuous innovation, leading to more robust, secure, and efficient BBS projects.

The rest of the paper is organized as follows. Section 2 provides a background for this study. We explain the related works in Section 3. The employed research method presented in Section 4 followed by the study findings in Section 5. Section 6 discusses the implications of this research. The threats of this research are presented in Sections 7. We summarized key learnings in Section 8.

## 2. Background

This section contextualises the proposed research in terms of background details and illustration of two fundamental concepts namely blockchain-based systems (in Section 2.1) and software engineering for blockchain (in Section 2.2). The concepts and terminologies introduced here, and illustrations provided in Figure 1, are used throughout the paper to elaborate technical details of SE for BBS.

### 2.1 Building and Operationalizing Blockchain Based Software

Blockchain-based software refers to any software-intensive systems, applications and/or services etc. that utilizes blockchain technology - enabled via distributed ledger mechanism - that enables secure and transparent recording of transactions across a network of computers [9]. BBS essentially consists of a few core elements such as a blocks, nodes, smart contract [1, 12, 13], as illustrated in Figure 1 (a). Central to BBS is a distributed ledger as a consensually shared database, synchronized across multiple locations in order to be accessible by multiple participants [6, 14] allowing the participants to independently authenticate, process, and validate transactions. Such verification can be done without the need for centralised authority or intermediary since every change made to the ledger would be reflected to all participants [6, 14]. BBS has a chain of blocks (i.e. records) that are sequentially linked together containing transactions. All transactions in the blockchain are recorded and included in the distributed ledger. In addition, the blocks have timestamp, hash values, and each block has a pointer referring to the data stored in the previous block in the chain [6, 11]. Also, a block saves an arbitrary set of transactional data that is created by a node (a computer in the distributed ledger network). At this point, each node on the chain stores a copy of the entire blockchain, and all nodes participating in the network can view and verify the entire chain. Hence, whenever a block gets an update (e.g., time stamp changed), a broadcast would be sent to all participated nodes in the distributed network by the creator node. The other nodes would perform validation for the new updates through predefined check (e.g., consensus mechanism). Once changes are validated, each node would add the block to their own local blockchain copy to provide a single source of truth. The validation process, including newly added block to the blockchain, is mainly performed by some nodes, called miners. In fact, miners get rewarded for successfully adding transactions to blockchain as an appreciation for using their computational power to mine for blocks [15]. Once all the participated nodes accept the transactions, recorded data in the block become non-reversible, transparent, and an immutable part of blockchain. These characteristics provide a secured mechanism for exchanging information between systems in a decentralized way, which make it possible to maintain records and tasks (e.g., financial transactions, medical records, etc.) [6]. These capabilities has the potential to increase the adoption for BBS. In fact, Casino et al. predicted that BBS will continue to be adopted in several domains such as manufacturing, cybersecurity, 5G networks, and Internet of Things (IoT) to enhance security by tracking records and ensuring the consistency of data [16].

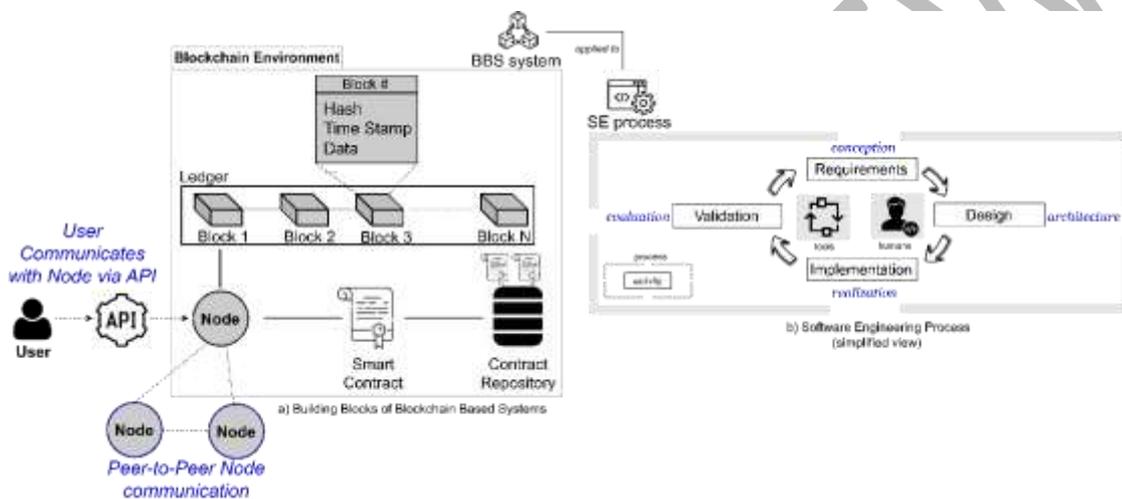

Figure 1. Fundamental Constituents of the Blockchain Architecture

## 2.2 Engineering Blockchain Based Software Systems (SE for BBS)

The software engineering process involves several themes that are mainly correlated, such as software requirements, process, testing, quality, maintenance, configuration management, and engineering management [17, 18]. Particularly, software engineering for BBS refers to a new type of software that uses the implementation of blockchain in its components [9]. Software development for BBS would revolve around using blockchain technology; and hence, can be perceived differently from non-BBS development. BBS, as highlighted in Section 2.1, consists of distributed ledger which forms the technical foundation for the development and use of innovative decentralised distributed systems. Furthermore, the development process of BBS includes process-centric engineering approach that encompass functional and quality aspects, design patterns and architectural models, source-code implementations, and testing, as depicted in Figure 1 (b). Engineering BBS remains challenging, due to the characteristics and limitations of such technology. These challenges raise the need for an in-depth research in this field in both technical and non-technical views, such as finding better approaches that balance between security and performance, supporting blockchain developers through providing better guidance, selecting consensus protocols in relation to transaction time and computing power requirements, and enhancing the cooperation of multiple institutions and stakeholders [3, 19, 20]. Such challenges and limitations can be linked to the immaturity of the technology itself, which is an obstacle to all new technologies when they first emerge. Nevertheless, other challenges could be directly linked to blockchain concepts and requirements. Conventional wisdom in software engineering confirms that the proper use of engineering methodologies ensures the development of quality software that satisfies the project's expectations while meeting the allocated budget and timeframe constraints [21]. Bearing in mind that once a system complexity increases, the more it becomes unmanageable and open to total breakdown [22]. Therefore, understanding and adopting a systematic engineering approach for BBS can reduce the possibility of user dissatisfaction, unanticipated security, and other defects that can be expensive to fix after BBS deployment. In light of common prudence in the field of software engineering, it will be challenging for a software team to analyse and identify the root causes of errors and get them fixed when things go wrong without a systematic approach [23-25]. In other words, the quality of development will greatly affect the quality of the resulting BBS running on the blockchain platforms [3, 9].

## 3. Related Work

In this section, we review the most relevant existing research that is categorized into three distinct areas, namely (a) empirical research on processes and practices to engineer BBS, (b) literature reviews on BBS engineering, and (c) solutions to design and develop BBS systems. We conclude the review with a comparative analysis between the existing relevant research and our study to justify the scope and contributions of our work.

### 3.1 Blockchain Software Engineering Processes and Practices

Numerous research efforts have been dedicated to exploring and recommending optimal processes and practices for blockchain software engineering. For instance, Chakraborty et al. [26] presented the first formal empirical study to explore software engineering practices in BBS projects, including requirement analysis, task assignment, testing, and verification. The authors identified 1,604 active BBS developers through Github repositories of 145 popular BBS projects and received 156 responses that met their criteria for analysis. The study findings revealed that code review and unit testing were the most effective software development practices among BBS developers. Interestingly, the requirements of the projects were mostly identified and selected through community discussion and project owners, differing from the requirement collection process in general open-source software (OSS) projects. Additionally, the study showed that development tasks in BBS projects were primarily assigned on a voluntary basis, a common practice in OSS projects. The results also highlighted the need to adapt standard software engineering methods, including testing and security best practices, to better address the unique challenges which BBS pose. Marchesi et al. [27] proposed a software development process for blockchain applications including requirements gathering, analysis, design, development, testing, and deployment. This process incorporates Agile practices, such as user stories and iterative incremental development, alongside more formal notations like UML diagrams with additions for blockchain-specific concepts. The study provided a detailed description of the method and includes an example to demonstrate its functionality, aiming to ensure software quality and adherence to software engineering principles in blockchain development.

Porru et al. [9] highlighted the necessity for software engineers to develop specialized frameworks and techniques for blockchain-oriented software development. The key factors for the future include effective testing activities, improved collaboration within large teams, and streamlined development of smart contracts. These factors are essential to ensure the successful advancement of blockchain-oriented software development. Destefanis et al. [28] highlighted the need for a discipline of Blockchain Software Engineering to address issues posed by smart contract programming and other applications running on blockchains. The study analyzed a case study involving a bug discovered in a smart contract library, which led to an attack on the Parity wallet application and the freezing of 500K Ethers (about 150M USD in November 2017). It examined the source code of Parity and the library, discussing how recognized best practices could mitigate such detrimental software misbehavior if adopted. The authors also emphasize the specificity of smart contract software development, which renders some existing approaches insufficient, and calls for the definition of a specific Blockchain Software Engineering discipline. Pinna et al. [29] presented an investigation into integrating agile blockchain-oriented software development principles with sustainability software design principles. Recognizing the importance of understanding the potential long-term effects of blockchain-oriented software use, the authors proposed a new Agile method for developing blockchain-oriented systems that incorporate sustainability awareness practices within the development phases, specifically in the requirements and acceptance tests. This approach enables addressing the sustainability of blockchain-oriented systems during the incremental and iterative development process. The paper described the process in detail and provided an example of its application in the supply chain sector.

### 3.2 Comprehensive literature reviews on BBS engineering

A range of literature review studies has been carried out, concentrating on various facets of BBS engineering practices, processes, tools, frameworks, and challenges. In this paper, we synthesize the pertinent literature reviews to complement and enrich our research findings.

Fahmideh et al. [30] conducted a systematic literature review on the state-of-the-art BBS engineering research from a software engineering perspective. The study focused on identifying the key aspects of BBS engineering, including theoretical foundations, processes, models, and roles. The study offers a comprehensive overview of development tasks, design principles, challenges, and resolution techniques, serving as a valuable resource for practitioners and researchers. This study establishes a foundation for future research in the field of BBS engineering. Demi et al. [31] conducted a systematic mapping study to provide an overview of blockchain-oriented software engineering applications. The authors identified 22 primary studies and extracted data on research type, topic, and contribution type. The findings revealed an increasing trend in research since 2018, suggesting blockchain's potential as an alternative to centralized systems, such as GitHub and Travis CI, and its ability to establish trust in collaborative software development. Moreover, smart contracts can automate various software engineering activities, including acceptance phases, payments, and compliance adherence. While the field is still maturing, this study offered valuable insights for researchers and practitioners interested in understanding how blockchain can transform the software development industry.

Vacca et al. [32] conducted a systematic literature review study focusing on software engineering challenges in developing smart contracts and blockchain applications. The authors examined six specific topics: smart contract testing, code analysis, metrics, security, Dapp performance, and blockchain applications. The paper reviewed various techniques, tools, and approaches proposed in the literature to address challenges posed by the development of blockchain-based software. In addition to the systematic review, the authors identified open challenges in each of the six topics that warrant further research, highlighting the need for continued exploration in this rapidly evolving domain. Lal et al. [33] conducted a comprehensive survey on the testing of Blockchain-based Applications (BC-Apps), discussing the challenges and state-of-the-art testing efforts for BC technologies. The authors first identifies challenges associated with BC-App testing and then provide an overview of existing testing tools and techniques for different components at various layers of the BC-App stack. The paper also proposed future research directions based on the identified challenges and gaps in the literature. The authors emphasized the need for standardized testing procedures and techniques for BC-Apps due to the unique nature of BC-based software development, which may render some existing tools or techniques inadequate. The study aimed to highlight the importance of BC-based software testing and promote disciplined, testable, and verifiable BC software development.

### 3.3 Proposed solutions for BBS engineering

A number of solutions have been proposed to enhance the effectiveness of BBS engineering activities. For example, Li et al. [34] proposed CrowdBC, a blockchain-based decentralized framework for crowdsourcing that addresses the limitations of traditional trust-based models, such as single points of failure and vulnerability to DDoS and Sybil attacks. By eliminating reliance on central servers or third-party institutions, CrowdBC ensures user privacy and requires only low transaction fees. The authors presented the architecture of the proposed framework, provided a concrete scheme, and implemented a software prototype on the Ethereum public test network using real-world data. Experimental results demonstrated the feasibility, usability, and scalability of the proposed system, showcasing its potential to enhance the development of crowdsourcing systems. Farooq et al. [35] proposed a blockchain-Based Software Process Improvement (BBSPI) approach to address the challenges faced by small and medium-sized enterprises (SMEs) in implementing software process improvement (SPI) initiatives. The authors introduced BBSPI and validate it through two case studies involving 55 representatives from 50 organizations. The results indicated that BBSPI can reduce SPI cost, time, and resource utilization while improving knowledge management and process maturity. This approach offered an efficient alternative to traditional centralized SPI, enabling SMEs to conform to common process improvement models while reducing costs, time, and resources.

Ulybyshev et al. [36] presented a blockchain-based solution for ensuring integrity, trust, immutability, and authenticity in cross-domain data communication and global software collaboration. The solution employs role-based and attribute-based access control to prevent unauthorized access and guarantee provenance of data integrity. It also detect data leakages by authorized blockchain network participants to unauthorized entities, supporting data forensics and provenance. Transactions within the global collaborative software development environment are recorded in the blockchain public ledger and can be verified at any time, with no possibility of repudiation. The authors proposed a modified transaction validation procedure to improve performance and protect permissioned IBM Hyperledger-based blockchains from DoS attacks caused by bursts of invalid transactions. Bose et al. [37] proposed an extensible framework based on standard provenance model specifications and blockchain technology for capturing, storing, exploring, and analyzing software provenance data in modern distributed software development environments. The proposed framework enhances the trustworthiness of provenance data, uncovers non-trivial insights through inferences and reasoning, and enables interactive visualization of provenance insights. The utility of this framework is demonstrated using open-source project data, showcasing its potential to improve comprehension, management, decision-making, and analysis of software quality, processes, people, and issues in heterogeneous software delivery contexts.

### 3.4 Comparative Analysis

We now present a summary of the most existing research relevant to our study, as presented in Table 1. Comparative analysis is based on four-point criteria including *(i) research challenge(s), (ii) focus and contributions, (iii) evaluation context, and (iv) research limitations.* The study reference points to an individual research work under discussion and its year of publication. Our work significantly contributes to the existing body of research in the domain of BBS engineering by presenting a process model that emphasizes development processes and methodologies specifically tailored for blockchain systems. The majority of related studies focus on the practices, challenges, and tools associated with BBS engineering; however, our research addresses the previously unexplored area of development processes in this context. By deriving a conceptual process model from the literature and validating it through a survey of 102 blockchain experts, we shed light on a crucial aspect of BBS engineering that has been largely overlooked in previous studies. Through a comparative analysis with related work, our research stands out as it delves deeper into the intricacies of development processes for blockchain systems, while also providing a comprehensive and validated process model. By doing so, we advance the understanding of BBS engineering and offer valuable insights to researchers and practitioners. Furthermore, our study bridges the knowledge gap within the literature and paves the way for future research in this rapidly evolving field, ultimately improving the overall quality and effectiveness of BBS engineering practices.

Table 1. Comparative Analysis of Most Relevant Existing Studies Compared to our Study.

| Study Ref. | Research Challenges | Focus and Contributions | Evaluation Context | Research Limitations | Pub. Year |
|---|---|---|---|---|---|
| **Literature survey** | | | | | |
| [32] | Reviewing the methods, techniques, and tools proposed for improving the design, construction, testing, maintenance, and quality of smart contracts and decentralized apps. | - Focusing on smart contract testing, security, code analysis and code metrics.<br>- Focusing on DApp performance measurement and blockchain applications.<br>- Stating the open challenges for the focus areas. | A total of **96 papers** were investigated in this literature review and identified the (previously mentioned) six focus areas. | Although the study introduced the common issues for each focus area, there are still some missing issues related to the design phase (e.g., permission design, skelton definition). | 2021 |
| [38] | Analysing the effects of current software engineering processes and models, such as Agile and DevOps, for better Software Process Improvement (SPI) in Blockchain-Oriented Software Engineering (BOSE). | - Exploring the necessity of integrating cutting-edge concepts and evolving existing software engineering methodologies for BBS.<br>- Discussing software project management practices within the context of BOSE development | A total of **24 papers** were selected for this study. | The study reviewed the existing software engineering processes and models (e.g., Agile, and DevOps) in BOSE; however, no further details were provided in each phase of the development life cycle which can be relevant to BBS. | 2022 |
| [18] | Investigated which types of blockchain technology would be advantageous for specific use cases in software engineering, emphasizing both their benefits and potential drawbacks. | - The study grouped the identified research topics into seven areas (i.e., *Software Requirements, SE Process, Testing, Quality, Maintenance, Configuration and deployment management,* and *Professional practices and guidelines*). | A total of **34 papers** have been considered in this study. | Although the study provided recommendations for the identified use cases, it is a bit shallow and requires in-depth analysis regarding the proposed blockchain SE use-cases. | 2022 |
| [31] | Enhance the understanding of blockchain-oriented software engineering by offering a comprehensive overview of software engineering applications facilitated by blockchain technology. | - Exploring the trend of studies utilizing blockchain in software engineering.<br>- Examining the various reported uses of blockchain as documented in the literature.<br>- Identifying the blockchain platforms employed in developing software engineering applications.<br>- Analysing how blockchain can contribute to the software engineering landscape. | A total of **22 papers** were selected for this study. | The study presented the findings regarding the uses of blockchain to support software engineering activities (i.e., requirements, software engineering process, testing, quality, maintenance, configuration management, software engineering management, and professional practice). | 2021 |
| **Surveys and case studies** | | | | | |
| [26] | Exploring software engineering practices in existing BBS projects. | Focusing on requirement analysis, task assignment, testing, and verification. | - Identified 1,604 active BBS developers through Github repositories of 145 popular BBS projects.<br>- 156 BBS developer **survey**. | - Survey questions were based on the proposed RQs in the study instead of having a solid literature review. | 2018 |

| Ref | Objective | Focus | Method/Evaluation | Limitations | Year |
|---|---|---|---|---|---|
| [27] | Proposing a software development process for blockchain applications. | Focusing on requirements gathering, analysis, design, development, testing, and application deployment as well as ensuring software quality and adherence. | - Incorporated Agile practices, such as user stories and iterative incremental development, alongside more formal notations like UML diagrams with additions for blockchain-specific concepts.<br>- The study evaluated the proposed development process through **developing a simple version of a voting system**. | - The study did not include practitioners' views to evaluate the proposed software development process for blockchain applications.<br>- Evaluation process through developing a voting system that might not be applicable to other domains. | 2018 |
| [9] | Highlighting the necessity for software engineers to develop specialized frameworks and techniques for blockchain-oriented software development. | Emphasizing the key factors for the future blockchain-oriented software development including effective testing activities, improved collaboration within large teams, and streamlined development of smart contracts. | - The study examined **193 repositories.**<br>- The study **extracted information on popularity** (Stargazers), programming languages, community involvement (Contributors, Open Issues, Watchers, Forks), and age (time elapsed since creation). | - The study focused on extracting information from existing BBS projects without including practitioners' feedback.<br>- The study did not provide an evaluation method for the emphasized key factors. | 2017 |
| [28] | Highlighting the need for a discipline of Blockchain Software Engineering to address issues posed by smart contract programming and other applications running on blockchains. | The authors emphasized the specificity of smart contract software development, which renders some existing approaches insufficient, and calls for the definition of a specific Blockchain Software Engineering discipline. It examined the source code of Parity and the library, discussing how recognized best practices could mitigate such detrimental software misbehavior if adopted. | - The study analyzed **a case study** involving a bug discovered in a smart contract library, which led to an attack on the Parity wallet application and the freezing of 500K Ethers. | - The study examined a coding issue (bug) rather than examining the development process of BBS with practitioners.<br>- The study proposed three main areas that need to be addressed (i.e., best practices and development methodology, Design patterns, and Testing) which require further evaluation. | 2018 |
| [35] | Proposing a blockchain-Based Software Process Improvement (BBSPI) approach to address the challenges faced by small and medium-sized enterprises (SMEs) in implementing software process improvement (SPI) initiatives. | The results indicated that BBSPI can reduce SPI cost, time, and resource utilization while improving knowledge management and process maturity. | - The authors, firstly, introduced BBSPI through **exploratory case study** followed by obtaining feedback from industrial experts.<br>- The study, then, validated the proposed BBSPI through **two case studies** involving 55 representatives from 50 organizations. | - The result might be biased due to the organization size.<br>- The results might be biased due to the limited number of participants (i.e., 50 participants). | 2021 |
| Proposed Study | Understanding the different aspects related to the development process for BBS. | Aiming to help researchers and practitioners in their quests on challenges and recommendations associated with the development of BBS systems | - Conducted a systematic literature review of 58 studies to identify a process, comprising of 23 tasks, to develop BBS systems.<br>- Engaging 102 blockchain experts in a survey to validate the BBS system development process. | - Solid literature review to derive 23 tasks to develop BBS.<br>- Diversity of data collection source to validate the identified tasks. | NA |

# 4. Research Design

In this section, we detail a four phase research design we conduct and document the proposed study. Each of the research phases is detailed below and illustrated in the Figure 2 below.

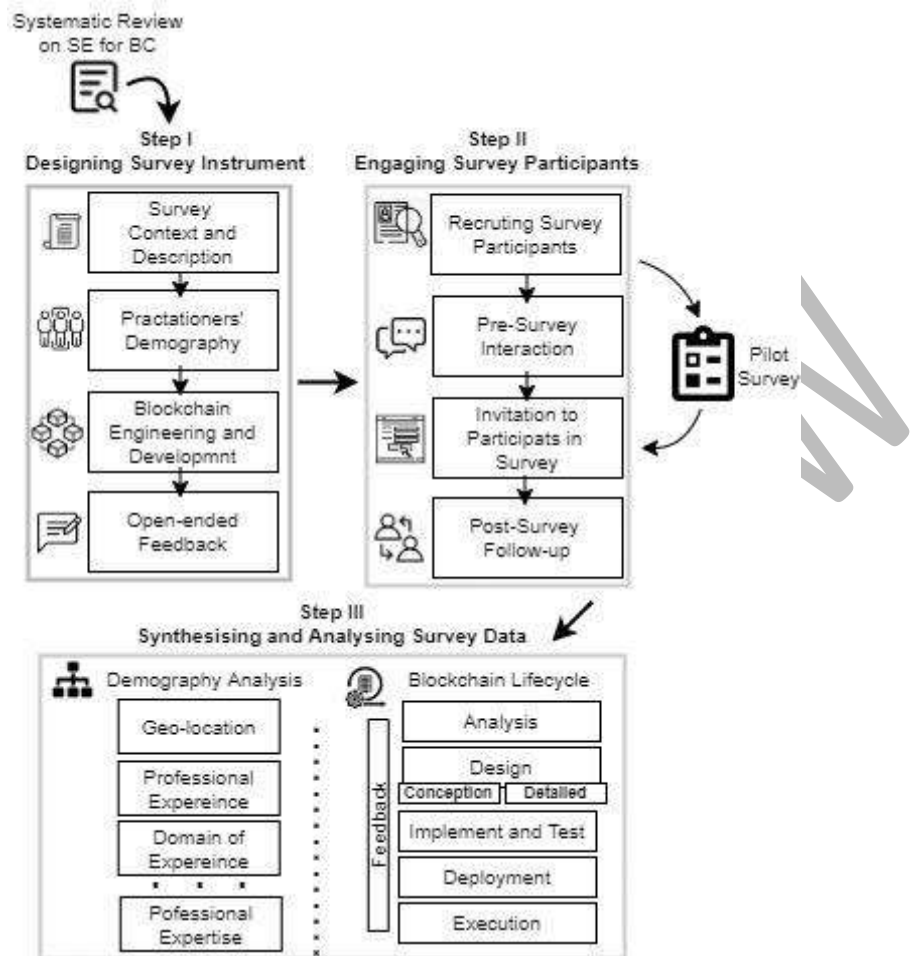

Figure 2. Overview of Research Method Adopted in this Study

## 4.1 Phase I – Designing Survey Instrument

The survey instrument for this research is based on our findings from a review study that has been conducted recently [6]. Bearing in mind the guidelines by [39] for conducting a survey-based studies which we also followed. We designed an online survey using Google platform which is easy to share, view, and manage across platforms. We briefly described the purpose and eligibility of our study in the survey preamble. As presented in Appendix 1, we divided our questionnaire into three main parts namely: (i) demographic questions (**Q1** – **Q9**), (ii) blockchain development process tasks (**Q1** – **Q31**), and (iii) open-ended question (**Q1** – **Q3**) to allow participants to share their experience when developing blockchain-based systems. For the illustration purpose, we briefly described the development tasks along with their definitions for our participants. We followed the description by asking the participants to rate the importance of each given task based on seven Likert-scale from 1 to 7, where 1 and 7, respectively, indicated not at all important and extremely important. The participants were also asked to suggest any missing important tasks and provide their reasons why these tasks are important and should be considered.

*Conducting pilot survey:* We conducted a pilot study for our survey to check if it was as coherent and concise as possible whilst gathering sufficient feedback. Minor issues reported and addressed were related to spellings, some terminologies, the sequence of questions and activities, and adding one question to the survey. The survey was designed in English and it took around 20 - 25 minutes to complete. This study has been approved by the Human Research Ethics Committee at the University of Wollongong with approval number 2020/423.

## 4.2 Phase II - Engaging Survey Participants

The survey was available online between July 2022 and September 2022 [40]. We were concerned about finding practitioners who can be qualified to be part of our study for two main reasons. The blockchain domain is a nascent area, and the entire population of blockchain domain experts was unknown to draw a sample size. Thus, we had to perform purposeful sampling [41] by manually checking the potential participants' profiles to ensure that they had real-world experience with the blockchain-based development process. We sought potential respondents in different online communities, in particular, LinkedIn, Twitter, Facebook, GitHub, and academic research groups. For instance, we used GitHub to mine the developers in blockchain repositories and to identify the email addresses of active developers, who had submitted changes and contributed to at least one blockchain project. After checking her/his online profile and confirming the credibility of each selected participant, we directly asked her/him to contribute and share their input with us by filling out an online survey. In total, we sent our study invitation to more than 740 randomly identified and verified blockchain experts through personalized emails or direct messages. Eventually, we received 102 valid responses (i.e. indicated by **P1** to **P102**).

## 4.3 Phase III – Demography Analysis of Survey Respondents

As indicated in Section 4.2, one of the recruitment methods was posting the survey on social media (e.g., blockchain-related groups on Facebook, and LinkedIn). Thus, it was not possible to calculate an accurate response rate for our survey. The demographic details (e.g., geographical distribution, participants' roles, years of experience, etc.). We were able to collect our data from 36 countries - six continents - in terms of their geographical distribution, as in Figure 3. The majority of the participants were from Australia and India (14.3%, 14 participants each). All our participants were involved in blockchain-related projects. In fact, they have been working in various blockchain application domains (e.g., retail, healthcare, logistics and transportation, manufacturing, education, smart energy, mission-critical systems, utilities, etc.). Our participants held various roles during the blockchain development process including but not limited to software developers, solution architects, core blockchain developers, smart contract developers, project managers, consultants, quality assurance engineers, and researchers. 18.4% of participants had 0-2 years, 29.6% had 3-5 years, 27.5% had 6-10 years, and 28.6% had over 10 years of general software development experience. For blockchain-specific experience, 53% had 0-2 years, 40.8% had 3-5 years, 9.2% had 6-10 years, and 1% had over 10 years. Additionally, 64.3% worked in teams of 1-10 members, 15.3% in 11-20 member teams, 16.3% in 21-30 member teams, and 8.2% in teams larger than 51 members. Additionally, we provided a multi-selection question to find out the different software development methodology to build BBS which our participants follow. Table 2 presents a detailed explanation for our participants selections.

Table 2. The Different Software Development Methodology to Build BBS which our Participants Used

| Categories for BBS development methodology | Number of participant and used methodology |
|---|---|
| **Category (i)**<br>*Participants who reported one methodology (n =64)* | - 20 participants do not follow any specific method.<br>- 17 participants used in house method, e.g., a combinational of methods.<br>- 17 participants followed agile methods, e.g., Scrum, XP, DSDM.<br>- 10 participants used object-oriented methods, e.g., Rational Unified Process (RUP) |
| **Category (ii)**<br>*Participants who reported more than one methodology (n =24)* | - 9 participants used in house method, and agile methods.<br>- 6 participants reported using object-oriented methods and agile methods.<br>- 3 participants combined agile methods, object-oriented methods, and in house method.<br>- 3 participants combined object-oriented methods and in house method.<br>- 1 participant combined Agile and no specific method.<br>- 1 participant combined in house method and no specific method.<br>- 1 participant combined object-oriented methods and no specific method. |
| **Category (iii)**<br>*Participants who reported a methodology that we did not provide within the given options (n =14)* | - 4 participants reported using continuous delivery and deployment (DevOps).<br>- 4 participants followed model-driven development.<br>- 4 participants used prototype development (Rapid Prototype).<br>- 1 participant used an actor model approach.<br>- 1 participant used the DAOM Methodology. |

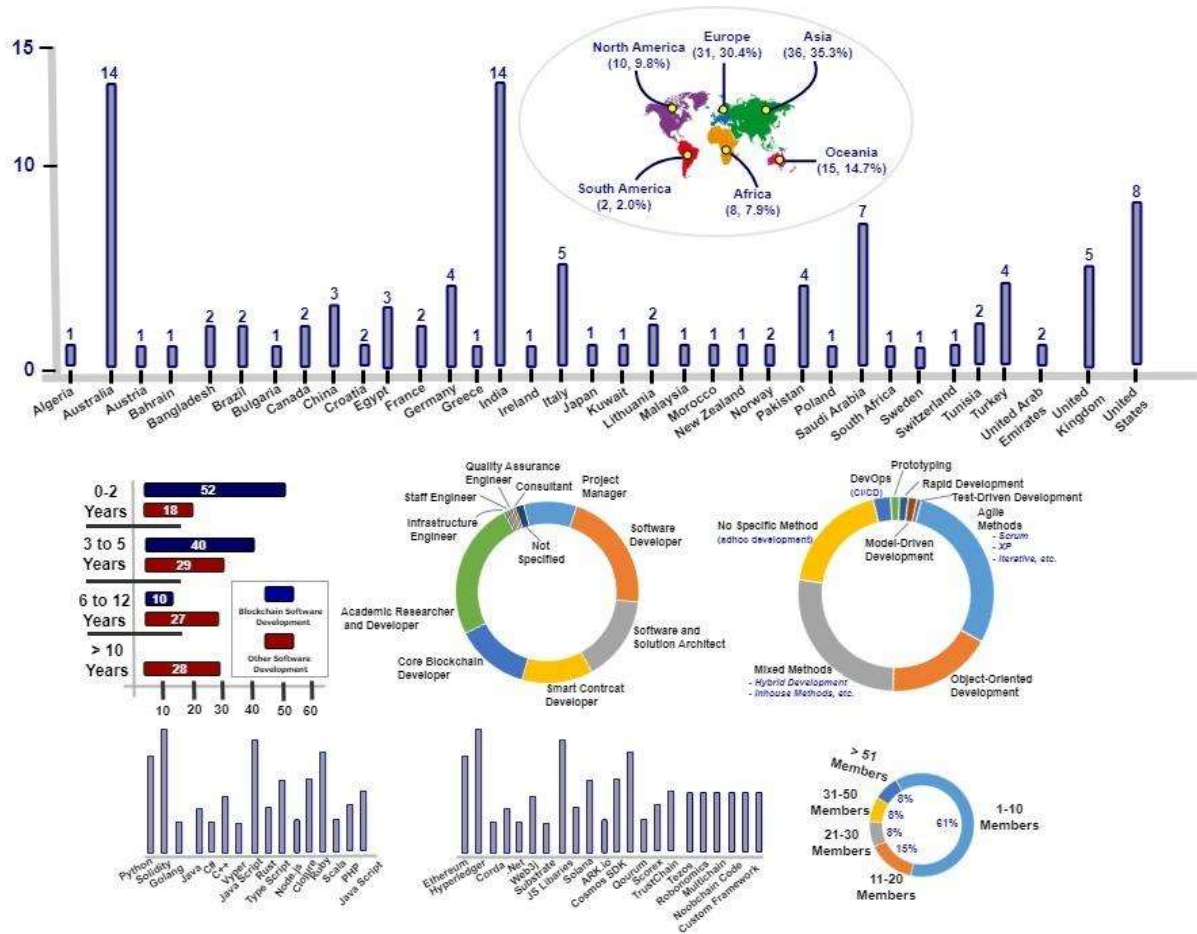

Figure 3. Demographic Details of our Study Participants (Sample Size =102)

**4.4 Phase IV – Synthesising and Analysing Survey Data**

Since we collected quantitative and qualitative data for this study, we followed three approaches to analyse data. Hereafter, we provide further details for each approach.

*Descriptive Analysis:* This approach helped us to present the demographic data (e.g., years of experience in the blockchain domain, development team size, etc.) for our study participants, as in Figure 3.

*Quantitative data analysis:* To understand the perceived importance of the different tasks when developing BBS among the different demographic groups, we used SPSS, a popular data analysis software. For each task in our framework (can be survey or questionnaire), we defined the null and alternative hypothesizes: *H0: the mean of the importance of the task is less than 5 vs. H1: the mean of the importance of the task is more than 5*. Hence, the task is perceived as important when it rated to or more than mid-point 5, i.e., somewhat important in our Likert-scale. Doing so has helped us to understand if the development tasks are distributed away from the median 5. This test was suitable for determining if the mean of an independent variable sample data is different from a specific value. It should be noted that **Kolmogorov-Smirnov** normality test was performed for the obtained responses, considering tasks as the dependent variables, to check the assumptions of statistical tests used in this study. The result ($p < 0.05$) suggested that the data were skewed (i.e., violated the normality assumption). Hence, we supported the usage of the statistical tests with the concept of central limit theorem [42], indicating that a sufficiently large (i.e., 102 in our case), has an approximate normal distribution. Further examination was conducted on the normal Q-Q plots generated by Kolmogorov–Smirnov test which showed that all the data points were close to the diagonal line. In addition, we considered the argument, has been verified by simulation results, that parametric tests (e.g., One-Way ANOVA Test) are not sensitive to the non-normality assumption in general [35], [36]. Lastly, we ensured meeting the T- Test assumptions in regard to collecting responses independently as none of the respondents were aware of the identity of other participants in this survey.

*Power analysis:* A post-hoc analysis was performed via the G*power 3 software [43] to check the statistical power of our performed **One-Sample T-Test**. The analysis takes the parameters effect size, i.e., d, sample size, and α error probability. Cohen [44] defines three levels to conceptualize the power, i.e., small effect ($d = 0.2$), moderate effect (d = 0.5), and large effect (d = 0.8). In this research, the alpha = 0.05, effect size d = 0.2, and sample size 102, were sufficient enough to achieve the statistical power 0.5 (1-β error probability).

**Qualitative data analysis:** As highlighted in Section 4.1, we included some open-ended questions to allow our participants to share their views in regards to the tasks that can be perceived importantly when developing BBS. We used **thematic analysis** method [45-47] to analyse the textual responses since it supports extracting the data and synthesizing the results. We used NVivo[1] software, a popular computer-based tool, to organize and analyze data. Coding was initially done by one of authors in the team that was reviewed and revised (wherever required) by another author to avoid potential bias.

## 5. Findings

### 5.1 Practitioners' perception about process tasks

**Perceived importance of task.** Based on the One-Sample T-Test results, shown on Table 3, we found that there is a statistically significant difference (*p-value <.001*) among our participants when they rated the importance of the majority of BBS tasks (i.e., 24 out of 26 tasks). Only two tasks were not statistically significant compared to other tasks. The two tasks which received less importance are: (1) incentive protocol design (*t-test statistic of =2,649, p-value of =0,009*), and (2) granularity design (*t-test statistic of=0,172, p-value of =0,864*). All our framework tasks were rated high importance, testing was rated as the highest task (*mean= 6,45*) and granularity design was rated as the lowest task (*mean=5,02*) by our participants. We also noticed that tasks such *requirement analysis* (*mean= 6,16*), *coding* (*mean= 6,14*), *and authentication and authorization design* (*mean= 6,06*) were received high rating compared with other tasks. These high ratings for the above mentioned tasks indicated that higher importance of testing during the development process of BBS, and higher importance of analysing the requirements when initiating BBS project. Regarding the different phases of BBS, the averages of rates for each phases of analysis, design, implementation, deployment, and maintenance are 5.73, 5.56, 6.30, 5.54 and 5.80, respectively. These averages of rates demonstrate that our participants believe implementation phase has more importance, followed by maintenance phase, followed by analysis phase, followed by design phase, and then the deployment phase.

Table 3. Descriptive Statistics for the Framework Tasks and the Results of One Sample T-Test for each Task

| Phase | Tasks | Mean | Std. deviation | T-Statistic value | p-value |
|---|---|---|---|---|---|
| Analysis Phase | 1. Feasibility assessment | 5,76 | 1,145 | 6,747 | <,001 |
| | 2. Requirement Analysis | 6,16 | 0,931 | 12,553 | <,001 |
| | 3. Actor identification | 5,50 | 1,041 | 4,850 | <,001 |
| | 4. Agreement Approval | 5,49 | 1,110 | 4,392 | <,001 |
| Design Phase | 5. Off/on blockchains identification | 5,67 | 1,011 | 6,692 | <,001 |
| | 6. Blockchain type selection | 6,05 | ,948 | 11,177 | <,001 |
| | 7. Permission design | 5,97 | 1,103 | 8,886 | <,001 |
| | 8. Skelton defination | 5,39 | ,997 | 3,974 | <,001 |
| | 9. State management design | 5,52 | ,982 | 5,342 | <,001 |
| | 10. Replication and synchronization design | 5,58 | 1,138 | 5,132 | <,001 |
| | 11. Authentication and authorization design | 6,06 | 1,205 | 8,800 | <,001 |
| | 12. Interaction design | 5,75 | ,886 | 8,489 | <,001 |
| | 13. Consensus protocol design | 5,72 | 1,084 | 6,666 | <,001 |
| | 14. Incentive protocol design | 5,30 | 1,159 | 2,649 | 0,009 |
| | 15. Granularity design | 5,02 | 1,152 | 0,172 | 0,864 |
| | 16. Dispute resolution design | 5,72 | 1,084 | 3,784 | <,001 |
| | 17. Gas consumption design | 5,30 | 1,159 | 5,385 | <,001 |
| | 18. Security design | 5,02 | 1,152 | 12,833 | <,001 |
| | 19. Change management | 5,35 | 1,369 | 2,604 | ,011 |
| Implementation Phase | 20. Coding | 6,14 | ,872 | 13,121 | <,001 |
| | 21. Testing | 6,45 | ,828 | 17,697 | <,001 |
| Deployment Phase | 22. Platform Configuration | 5,54 | ,961 | 5,665 | <,001 |
| | 23. Publishing | 5,54 | ,982 | 5,547 | <,001 |
| Execution and Maintenance Phase | 24. Execute | 5,80 | 1,096 | 7,356 | <,001 |
| | 25. Finalize | 5,78 | 1,006 | 7,814 | <,001 |
| | 26. Monitor | 5,81 | 1,056 | 7,730 | <,001 |

---
[1] https://www.qsrinternational.com/nvivo-qualitative-data-analysis-software/home/

**5.2 Demographic analysis.**

In order to understand the demographic differences in the perceived importance of the given tasks. We tested the null hypothesis (i.e., $H_0$: *there is no significant difference*) against the alternative hypothesis (i.e., $H_1$: *there is a significant difference*). We performed One-Way ANOVA test to check if the dependent variables (i.e., tasks) would be different by the demographic independent variables (i.e., years of experience in developing BBS). Out of 26 development tasks, we observed that there are statistically significant differences for four tasks, namely *replication and synchronization design* ($F (2, 99) = 5.48$, p=.006), *authentication and authorization design* ($F (2, 97) = 5.64$, p=.005), *gas consumption design* ($F (2, 99) = 6.76$, p=.002), and *testing* ($F (2, 99) = 8.82$, p=.000).

A Tukey post-hoc test was performed for the tasks which we found significant to further understand where the differences within the three groups lie. We found that there are significant differences among the three groups when perceiving the importance of the task *replication and synchronization design* (p = .004 between 0 - 2 years of experience and 6 - 10 years of experience, p = .011 between 3 - 5 years of experience and 6 - 10 years of experience). Participants with 0 - 2 years of experience perceived higher importance for *replication and synchronization design*, mean rank = 5.73 which is less than the mean rank = 5.65 for 3 - 5 years of experience and less than 6 - 10 years of experience (mean rank = 4.50).

The results also showed that the importance of task *authentication and authorization design* was only significantly different between the participants who had 0 - 2 years of experience compared to those with 6 - 10 years of experience (p = .005). In other words, the importance of the task *authentication and authorization design* was perceived higher from those who had 0 - 2 years of experience (mean rank = 6.18), which is less than participants who had 6 - 10 years of experience (mean rank = 4.90).

For the *gas consumption design* task, we noticed significant differences among the three groups when perceiving its importance (p = .001 between 0 - 2 years of experience and 6 - 10 years of experience, p = .003 between 3 - 5 years of experience and 6 - 10 years of experience). Our statistical results indicated that participants with 0 - 2 years of experience perceived higher importance for *gas consumption design task*, mean rank = 5.79 which is less than the mean rank = 5.73 for 3 - 5 years of experience and less than 6 - 10 years of experience (mean rank = 4.40).

We also examined the differences within the three groups in regard to the *testing* task. Our results revealed significant differences among the three groups when perceiving the importance of the task testing (p = .000 between 0 - 2 years of experience and 6 - 10 years of experience, p = .002 between 3 - 5 years of experience and 6 - 10 years of experience). In fact, we noticed that participants with 0 - 2 years of experience perceived higher importance for *testing* task, mean rank = 6.62 which is less than the mean rank = 6.48 for 3 - 5 years of experience and less than 6 - 10 years of experience (mean rank = 5.50). Therefore, we reject $H_0$ and accept $H_1$ concluding that participants' perceived importance of the development tasks (*replication and synchronization design, authentication and authorization design, gas consumption design, and testing*) for BBS development years differently based upon their years of experience.

Table 4. Univariate Analysis between the Demographic Variables (independent variables) and the Perceived Importance Tasks, i.e., task rating (dependent variables)

| Task | Years of experience in developing BBS (One-Way-ANOVA Test) | | | | | | F | p-value |
|---|---|---|---|---|---|---|---|---|
| | 0 - 2 years (N=52) | | 3 - 5 years (N=40) | | 6 - 10 years (N=10) | | | |
| | Mean | Std. deviation | Mean | Std. deviation | Mean | Std. deviation | | |
| 1. Feasibility assessment | 5.67 | 1.115 | 5.85 | 1.027 | 5.90 | 1.729 | .343 | .711 |
| 2. Requirement Analysis | 6.17 | .944 | 6.15 | .975 | 6.10 | .738 | .027 | .973 |
| 3. Actor identification | 5.48 | .939 | 5.48 | 1.154 | 5.70 | 1.160 | .202 | .818 |
| 4. Agreement Approval | 5.58 | 1.109 | 5.38 | 1.115 | 5.40 | 1.174 | .362 | .697 |
| 5. Off/on blockchains identification | 5.61 | 1.002 | 5.78 | .947 | 5.60 | 1.350 | .331 | .719 |
| 6. Blockchain type selection | 6.12 | .832 | 6.05 | .986 | 5.70 | 1.337 | .802 | .451 |
| 7. Permission design | 6.19 | .930 | 5.75 | 1.296 | 5.70 | .949 | 2.202 | .116 |
| 8. Skelton defination | 5.48 | 1.038 | 5.28 | .933 | 5.40 | 1.075 | .477 | .622 |
| 9. State management design | 5.50 | .897 | 5.50 | 1.155 | 5.70 | .675 | .184 | .832 |
| 10. Replication and synchronization design | 5.73 | .866 | 5.65 | 1.210 | 4.50 | 1.581 | 5.479 | .006 |
| 11. Authentication and authorization design | 6.18 | 1.207 | 6.20 | .992 | 4.90 | 1.449 | 5.638 | .005 |
| 12. Interaction design | 5.79 | .825 | 5.75 | .899 | 5.50 | 1.179 | .440 | .645 |
| 13. Consensus protocol design | 5.77 | 1.002 | 5.80 | 1.018 | 5.10 | 1.595 | 1.826 | .166 |
| 14. Incentive protocol design | 5.44 | 1.018 | 5.20 | 1.181 | 5.00 | 1.700 | .874 | .421 |
| 15. Granularity design | 5.17 | 1.098 | 4.75 | 1.256 | 5.30 | .823 | 1.887 | .157 |
| 16. Dispute resolution design | 5.46 | .979 | 5.33 | 1.047 | 5.10 | .876 | .625 | .537 |
| 17. Gas consumption design | 5.79 | 1.091 | 5.73 | .905 | 4.40 | 1.838 | 6.756 | .002 |
| 18. Security design | 6.46 | .896 | 6.35 | 1.331 | 6.00 | .816 | .777 | .463 |
| 19. Change management | 5.52 | 1.336 | 5.28 | 1.450 | 4.80 | 1.135 | 1.271 | .285 |
| 20. Coding | 6.13 | .886 | 6.26 | .880 | 5.70 | .675 | 1.642 | .199 |
| 21. Testing | 6.62 | .661 | 6.48 | .784 | 5.50 | 1.179 | 8.818 | .000 |
| 22. Platform Configuration | 5.52 | .939 | 5.58 | .931 | 5.50 | 1.269 | .046 | .955 |
| 23. Publishing | 5.52 | 1.019 | 5.53 | .987 | 5.70 | .823 | .147 | .864 |
| 24. Execute | 5.92 | .882 | 5.69 | 1.341 | 5.60 | 1.075 | .679 | .510 |
| 25. Finalize | 5.85 | .849 | 5.78 | 1.143 | 5.44 | 1.236 | .609 | .546 |
| 26. Monitor | 6.00 | .886 | 5.60 | 1.277 | 5.67 | .707 | 1.742 | .181 |

### 5.3 Challenging process tasks (Other important tasks) in practice

In addition to the close-ended questions, which we asked participants to rate the importance of the predefined tasks for the development process of BBS (illustrated in Figure 4 below), we provided open-ended questions (presented at the end of each phase of the development process) to allow participants share 'other' tasks, detailed below, they consider as important but were not presented within the survey questions. The presented statements capture the input of 102 participants (referred to as **P1** to **P102**).

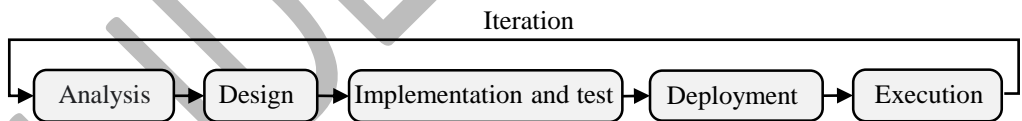

Figure 4. An Overall Development Process for BBS

### A. Challenges (Important tasks) related to analysis phase:

The analysis phase defines the requirements of a BBS and establishes a link between stakeholders' needs and desired features. 29 participants highlighted other tasks, such as platform selection, performance, cost, security, etc., that can be considered as an important during the analysis phase of BBS. In fact, two participants (i.e., **P4**, and **P59**) emphasized that analysing phase should be done to understand the needs, and make the remaining phases easy and faster. **P4** stated

*"analysing the need for a blockchain in the first place. Many times Blockchain hinders usability and ease of use, which turn the blockchain technology as a burden rather than a solution."*

**P59** has also addressed this concern by mentioning

*"With the problem well defined the way to solve it becomes simpler, a blockchain does not always need high speed or smart-contracts"*

Cost analysis is a critical aspect of BBS projects as it helps to manage the available resources, manage the associated risks, and foremost scalability in the future. Analysing the cost has been reported by a few participants (**P6, P84,** and **P86**). **P6** and **P86** respectively stated

> *"analyzing business requirements and business value most important for analyzing blockcahin system"*
>
> *"Future cost of transactions to be projected upfront"*.

Security and privacy concerns are essential practices for protect information and maintaining the integrity of transactions in BBS. In fact, performing security analysis, ensuring privacy protection, and conducting risk assessment for the BBS project have been also perceived as important tasks to be considered as **P24, P30, P35, P60, P68, P84,** and **P91**. For instance, **P84** wrote

> *"System Security Consideration (e.g., how to protect the private key ?)"*.

Additionally, conducting a risk assessment in BBS is essential to identify and mitigate potential vulnerabilities, **P68** stated

> *"In the design phase, the risk assessment and mitigation can be considered, which should be included in the backward arrow to the analysis phase"*.

Analysing the selection of the potential BBS platform is a critical step that ensures the selected platform aligns with the project's goals (e.g., technical, security, and scalability requirements, etc.) Analysing the selection of the platform for BBS project has been also perceived as an important task by six participants (**P11, P23, P36, P45, P60, P76** and **P101**). **P11** and **P76** respectively wrote

> "*analysis of the blockchain platform for deployment and execution*"
>
> *"Blockchain platform selection as a separate task, when a decision was made to utilize a public/private blockchain, could also be included. As some restrictions impacting the design of the system components are introduced by some specific platform selections."*

Indeed, two participants (**P78**, and **P91**) reported that platform selection should be depending on the business use case. **P91** stated

> *"Outline plan must consider the type of blockchain depending on use-case as well."*

One participant (**P83**) reported that ensuring stakeholders readiness to shift to blockchain technology (BT) is a crucial task in the analysis phase. **P83** wrote

> *"In the analysis phase is really crucial knowing the readiness (for using BT) of the stakeholders involved in the BT-based projects."*

It should be noted that there are other tasks that have been highlighted by our participants such as analysing performance and scalability (**P20**, **P32**, and **P82**), contract and transactions analysis (i.e., **P46**, and **P60**), analysing usability and ease of use (**P4**), analysing the architecture (**P8**), domain analysis (**P75**), analysing governance and policy to be adhered (**P69**).

### B. Challenges (Important tasks) related to design phase

The design phase is mainly exploit and map the defined outcomes produced by the analysis phase and transform them into designing BBS. Whilst the majority of participants found the tasks, which we presented in the questionnaire comprehensive for all design tasks (e.g., **P31** wrote *"...these are all important tasks and they depend on the specifics of the system under-development..."*), 18 participants have shared their thoughts in regard to the design phase and highlighted some tasks (e.g., smart contracts and components design, architecture design, transaction design, etc.) that need to be considered in this phase.

Designing smart contracts aimed to (i) analyze existing textual legal contracts, (ii) identify and convert semantic clauses and rules that are expressed as smart contracts, and (iii) transform smart contracts expressible as software codes. Designing smart contracts and components were perceived as an important task to be considered by three participants (**P6, P11,** and **P45**). **P6** and **P11** respectively stated

> "*design of smart contracts and blockchain components*"
>
> "*Design of services and components that can be mapped to the data blocks in the chain*".

In BBS, transaction design helps to specify the entry points at which function codes of a smart contract should be executed. The transaction design involves many aspect such as, transaction fee, transaction performance, transaction verification, and transaction constraints etc., that need to be also considered during the design of BBS. Three participants (**P14, P32, P36**) emphasized that quality and transaction design is an important task in this phase. **P14** wrote *"quality design (transaction processing time, block mining)"*, **P32**, and **P36** both wrote *"Transaction Design"*.

Authentication and authorization design for BBS, which refers to specific mechanisms to enable the identity detection of an entity in a blockchain network, has been highlighted by two participants (**P54**, **P92**). **P54** suggested using public key system for authentication,

> *"The usage of public key system like ED25519. Although majority of blockchains use this for SSH authentication, there are other options, as well as using a customized version of a present zero-knowledge proof systems suitable for the blockchain's purpose"*

**P92** indicated that authentication can be an off-chain option and more focus to be paid on authorisation, **P92** wrote

> "*authorisation is critical, but authentication is often not (or standard/off-chain)*".

One participant (**P86**) mentioned that the design of BBS should reduce gas fee for some blockchain types. The participant wrote

> *"Gas fee needs to be 0, as per design or else power consumption per transaction will harm environment"*.

There are other tasks which has been highlighted by other participants such as application and interface design (**P35**, **P80**), blockchain design patterns (**P8**) the overall architectural design (**P101**).

## C. Challenges (Important tasks) related to implementation phase

The implementation phase is more concerned with coding and testing BBS. While **P86** found this phase as time consuming *"This is most time consuming step so you will need a coder and mathematician both"*, 17 participants have shared their thoughts in regard to the implementation phase and suggested other tasks that need to be considered. BBS can integrate off-blockchain and on-blockchain components. Hence, code refactoring is the process of restructuring code, while not changing its original functionality and ensuring interoperability requirements, seeking approvals of component changes, and ensuring the entire BBS functions correctly. Code refactoring has been perceived as an important task during BBS by three participants (**P11**, **P22**, **P36**). For instance, **P22** wrote *"Exception handling, code refactoring."* And **P36** wrote *"code maintenance and refactor"*.

Using simulator help to test the effectiveness as well as identifying the strengths and weaknesses. Three participants (**P6**, **P32**, **P60**) recommended conducting a simulation testing in the implementation phase. **P6** wrote

> *"simulation of smart contract execution is important in blockchain testing."*

Testing for interoperability and automation have been perceived as an important task during BBS by four participants (**P6**, **P68**, **P91**, **P92**). In fact, **P68** assured that off-chain components should be checked for interoperability. **P68** wrote

> *"The off-chain components should also be tested, to ensure the interoperability of on-chain and off-chain components."*

Whilst BBS may raise complexities once changes occur, two participants (**P4**, **P59**) pointed out this issue and suggested considering a mechanism to overcome this concern. **P4** wrote

> *"... Also does the smart contract put in place a mechanism that considers later changes after being deployed?"*

**P59** suggested using a methodology to reduce the challenges in the long run. **P59** stated

> *"Using methodology like TDD, DDD, DevOps, good documentation, mainly good documentation, can minimize challenges in the long run."*

There are other tasks that has been indicated by individually. For example, **P33** suggested a third-party to perform auditing. **P33** wrote *"Audit phase: third party should audit the smart contracts"*. Other tasks that participants indicated are: Testing scalability, TBS, i.e., transactions per second, (**P91**), test case generation and execution (**P45**), keep logging of transactions (**P32**), and ensure implementing exceptions handling (**P22**).

### D. Challenges (Important tasks) related to deployment phase

Deployment phase is concerned about deploying the final developed BBS. We presented two important tasks (i.e., *platform configuration, and publishing*) for our participants to rate and asked them add, in case there are any other tasks that should be considered. **P33** mentioned that the presented tasks can be seen differently depending on the blockchain selection. **P33** stated

> "*These steps vary greatly depending on the blockchain we choose, if we go with some public blockchain (e.g. Ethereum) platform configuration step is out of our reach. If we deploy our own blockchain then it is important of course.*"

A few participants have suggested some tasks that can be considered within the deployment phase. For example, **P4** recommended conducting testing to ensure the implemented BBS satisfies the specified requirements from the analysis. **P4** wrote,

> "*Conducting tests after deployment and load testing as well. This would reveal a set of phenomena that would be discovered during the testing phase.*"

**P92** suggested to work on the security in the deployment phase as well as taking upgradeability into account. **P92** wrote

> "*Managing security of deployment and consider upgradability.*"

Other participants suggested tasks such as detailed analysis and selection of deployment environment (**P11**), continuous delivery (**P101**), and the versions of smart contracts (**P86**).

### E. Challenges (Important tasks) related to execution and maintenance phase

The maintenance phase in BBS aims to ensure that each components of the blockchain continues to function as it was designed, and perform the required repairs or upgrades once needed. For this phase, we presented three main tasks, i.e., *execute, finalise and monitor*, followed by an open-ended question to seek their thoughts in this regard. Monitoring the overall performance was suggested by **P11**,

> "*monitoring the performance of chain execution and transmit over network*".

The ability to validate earlier transactions has been perceived as an important task in this phase by two participants (**P8, P24**). **P24** and **P8** respectively wrote

> "*execution logging and tracing to be maintained*"

> "*Maintaining Execution logs*".

In addition, **P14** suggested ensure having the ability of block recovery, and transaction reversal. **P32** perceived *fault tolerance*, *maintenance and evolution of blockchains* as important tasks to be considered within the maintenance phase for BBS.

## 6. Discussion and Future Works

### 6.1 Understanding the development processes for BBS: key tasks and activities (RQ1)

Our study highlighted the key tasks for the development phases of BBS, and based on feedback from 102 participants. These tasks emphasize the complexity and multifaceted nature of BBS development. We discuss the findings form the statistical results and the key tasks which have been indicated by our participants.

It would be important to recognize the unique challenges when developing BBS which may exist at different levels. This indicates that not all traditional software engineering literature can be equally applicable within the BBS context. In BBS projects, there are different components (e.g., blocks, nodes, distributed ledgers) and different characteristics (e.g., large amounts of data, design interactions between software and hardware components, collaboration among various stakeholders, various user interfaces, and security and reliability issues) that software development teams have to deal with. For instance, compared to traditional software engineering, identifying rapidly-changing requirements from various key stakeholders in BBS is a difficult task. Among all the 26 tasks that we presented for our participants and based on the years of experience (Table 4), *testing task* was received the highest importance of all tasks for those who have 0 – 2 years of experience, the mean rank= 6.62, and the highest for those who have 3-5 years of experience, mean rank=6.48. Yet, *testing task* was not the highest rating among other tasks for those who have 6 – 10 years of experience (mean rank = 5.50). *Requirement Analysis* had the highest rating from those who have 6 – 10 years of experience (mean rank = 6.10). In addition, we noticed that participants who have 6 – 10 years of experience had less mean rank compared with other groups for the vast majority of our tasks. The only tasks that had higher rating were *actor identification* (mean rank = 5.70), *granularity design* (mean rank = 5.30), and *publishing* (mean rank = 5.70). Furthermore, 14 out 26 tasks that asked our participants to rate, we noticed that the mean ranks were in ascending order (i.e., participants who have 0 – 2 years of experience had higher rating, followed by participants with 3-5 years of experience, followed by those who have 6 – 10 years of experience). This lead us that the development tasks for BBS can be viewed differently based on the seniority level. This could have an indication that participants with less experience had more concerns about the ways of handling the development tasks, which can be slightly different than those who have a rich experience. Further investigation can be done to explore these aspects through conducting a semi-structured interviews that would allow to gather in-depth views.

By examining the important tasks and activities associated with each phase (Figure 4, i.e., analysis, design, implementation, deployment, and maintenance). The *analysis phase* requires a comprehensive understanding of requirements and stakeholder needs. Participants confirmed the importance of tasks like *platform selection*, *cost analysis*, *security analysis*, and *risk assessment*. For example, **P4** and **P59** emphasized understanding the need for blockchain to avoid unnecessary complexity. Other critical tasks include *performance* and *scalability analysis*, *contract* and *transaction analysis*, and *stakeholder readiness assessment*. These tasks suggest that the *analysis phase* must consider various technical, financial, and strategic factors to establish a robust foundation for BBS. In the *design phase*, participants agreed with the predefined tasks but also highlighted additional considerations. *Smart contract design* and *transaction design* were noted as critical, with **P6** and **P11** emphasizing the technical challenges involved. *Authentication and authorization mechanisms*, *gas fee reduction, application and interface design, and overall architectural design* were also highlighted. These tasks confirmed the need for detailed planning and consideration of various design elements to create an efficient BBS. Participants in the *implementation phase* noted the importance of *code refactoring, ensuring interoperability, and conducting simulation testing*. **P22** and **P36** emphasized maintaining and refactoring code to ensure functionality. Participants also highlighted the need for mechanisms to handle *post-deployment changes*, with suggestions for methodologies like DevOps. Other tasks identified include *third-party auditing, scalability testing, and maintaining transaction logs*. These insights highlight the iterative nature of the *implementation phase*, requiring *continuous testing and refinement*. In the *deployment phase*, participants indicated that tasks could vary significantly depending on the blockchain platform. **P33**'s comment confirmed the dependency of deployment tasks on the specific blockchain used. Additional tasks include *managing security*, and *ensuring upgradeability*. These tasks are critical to ensure the system operates as intended and can adapt to future requirements. Lastly, the *maintenance phase* ensures the BBS continues to function effectively. Participants emphasized *performance monitoring, execution logging, and the ability to validate* and *recover transactions*. **P32** highlighted the need for *fault tolerance* and *blockchain evolution*. These tasks emphasised the importance of ongoing maintenance and adaptability for BBS.

Based on the key tasks that have been raised by our participants, the following implications can be drawn (i) Analysis and Planning: A holistic approach to requirements gathering and planning is essential, considering technical, financial, security, and stakeholder readiness factors, (ii) Design and Testing: Thorough design and rigorous testing are crucial. Practitioners should plan and test all aspects, such as smart contracts, transaction processing, interoperability, etc, (iii) Adaptability and Scalability: Designing adaptable and scalable systems is vital. Ensuring that BBS can handle changes post-deployment is crucial for long-term success, and (vi) Cross-Phase Considerations: Decisions made in one phase can impact others. For example, platform selection in the analysis phase can influence design constraints and implementation challenges.

**6.2 Tailoring the existing methods for BBS Development (RQ2)**
It is generally accepted that a software engineering approach needs to be tailored to the project context, and hence any utilized method should be adjusted to the usage circumstances [48]. The work by Miraz and Ali [49] examined whether traditional software development lifecycles are appropriate to use in the context of BBS. The study concluded that software teams may omit particular aspects in an engineering approach not out of ignorance but rather out of a practical basis that those aspects are not pertinent to the functions of a chosen blockchain platform, the choice of blockchain type, or the project's setting itself. Our participants have several views in regards to tailoring or adjusting their in-house software development methodologies (e.g., Agile methods, SDLC, etc.) to fit for the development of BBS. For example, **P33** indicated *"It depends on what you work on within blockchain*

*ecosystem, but in, general blockchain development can be seen as any other software development so in my opinion any methodology can be applied."* **P32** who reported using Prototype development stated *"The nature of the solutions we develop requires rapid prototyping so prototypical development with transaction modeling and secure code development are more prominent in blockchain."* **P59** who reported using Actor models approach mentioned *"A change of mentality is needed, because depending on the project, updates and new functionalities will not be implemented immediately in the system, because it takes engagement from the miners."* Six participants (i.e., **P20**, **P24**, **P42**, **P50**, **P90**, and **P97**) reported using agile to ensure smooth development for BBS project. **P97** wrote *"We use an agile method that allows us the flexibility we need for adjusting our approach as we operate. We don't do anything specific to optimize for blockchain usage though."*, and **P90** stated *"Agile Methods and Coordinated in house methods are being used."*

From the above discussion, we imply the necessity for tailored software engineering approaches in BBS development. Our findings emphasize the importance of adapting traditional methodologies, embracing agile practices, and fostering flexibility to meet the unique demands of BBS projects. Additionally, the need for specialized development strategies, secure code practices, and a shift in development mentality highlights the complexity and evolving nature of blockchain technology. These insights provide a roadmap for practitioners to optimize their development processes for successful BBS implementation.

**6.3 Key differences between conventional software development and BBS Development (RQ3)**

It would be argumentative whether or not there are any differences between conventional software development methodologies (e.g., Agile methodologies, object-oriented methodologies, etc.) and the existing methodologies for developing BBS. We believe that development process of BBS can be well-connected and positioned with the conventional software engineering development process in the overall lifecycle. Nevertheless, it can be different in some other perspectives. Diving into the details during the development process of BBS would result in finding specific issues that can only exist in BBS context. For instance, BBS requires ensuring the security, reliability, and accepting the immutability of system transactions. BBS also deals with challenges related to smart contracts, different attributes for transactions (e.g., verification, fee, security, etc.), consensus mechanisms, gas consumption which cannot be applicable for other software development process. Only eight out of 102 participants did not think that there are distinguishing differences between the development process of BBS and conventional software development process. **P14** stated *" For me not much difference, as a software programmer I use APIs and libraries like blockstack and core chain and it seems like programming with immutable objects and linked lists"*, **P28** wrote *"My personal opinion: You don't have a difference. The end of day: Blockchain is a software program."*, and **P53** *"We are developing our blockchain products as any other softwares."*

The implications that can be drawn is the highlighted relationship between conventional software development methodologies and those tailored for BBS. While integration with traditional practices is feasible and beneficial for managing lifecycle phases, the unique challenges inherent to BBS necessitate specialized approaches. These challenges include ensuring transaction security, reliability, and managing blockchain-specific attributes like smart contracts, consensus mechanisms, and gas consumption. Participants' perspectives vary, with some viewing BBS development as analogous to traditional software programming, emphasizing the continuity in principles and tools. However, the majority acknowledged the need for tailored methodologies to effectively address blockchain's distinctive requirements. This recognition underscores the importance of continuous adaptation, specialized education, and collaborative innovation to optimize BBS development practices across diverse industries and regulatory landscapes.

**Implications for practice.** We consider the proposed framework within this study can be beneficial for those who are working on the development process of BBS based on the following views:

(i) Our framework can be held as a guideline for senior managers which can help to anticipate the different tasks and scenarios when developing BBS. According to the obtained responses, the presented framework contained a full list of tasks and recommendations that software engineers for BBS should bear in mind.

(ii) Our framework can be treated as a tool or to-do checklist by software teams to assess the extent to which their in-house development approach supports the development of BBS. In addition, the framework's tasks can be utilized within the adopted development approach to extend its capability for BBS development. Moreover, it can be used as an evaluation tool to identify shortcomings, strengths, similarities, and differences among alternative approaches for BBS development. Particularly, the framework makes it easier to select the appropriate approaches that suit situation-specific characteristics associated with a specific BBS project. This normative application of the provided framework is consistent with software process improvement as quality management activities are made by software teams to ensure high-quality, repeatable development processes in a cost-effective manner, in this case for BBS.

(iii) We believe that our framework can be applied to the system development process of other computing paradigms. Software development teams can view the identified tasks with the development process, and consider the results of this study, as can be seen in their development projects.

# 7. Threats to Validity

We now discuss some of the threats to the validity of this study. There are three types (i.e., internal, construct, and external) validity threats to be discussed below.

### 7.1 Internal Validity

Internal validity refers to the extent to which the observed results were from a reliable population. In our study, we ensured that we selected the participants who have the right kind of expertise to be part of our research (e.g., seeking potential participants and screening their publicly available profiles). We also explicitly described the nature of our research and the expected characteristics of the target participants within the survey preamble. We estimated the time to complete our survey, which can take 20 - 25 minutes. We knew the fact that participants were required time/effort and concentration to respond the questions accurately. The issue (i.e., fatigue bias) may lead some participants to complete the survey in a cursory way, which may have had a negative impact on the rating of the importance of BBS tasks or even for the open-ended questions; and thus, limited the validity of the reported statistical tests and quotes. We excluded odd responses and contacted participants to double-check the accuracy of her/his responses. Other threat could be relying on practitioners' survey for data collection and lack of triangulation which could have impacted the reliability of the findings. Another possible threat could be researchers' bias during data analysis, which could have been misinterpreted when analyzing the qualitative data. To overcome this threat, we performed (i) initial coding of the data followed by (ii) evaluation and finalization of the codes. Lastly, there can be a risk of social desirability bias (i.e., participants try to provide responses that a researcher want). We assured the participants that all the collected data would be kept anonymous to meet the human ethics approval obtained for this study.

### 7.2 Construct Validity

Construct validity refers to the extent to which the used instrument measures what it is supposed to be measured. Employing the suitable instrument for data collection and synthesis can threaten the validity of study results. To ensure un-biased data collection, we designed an initial version based on our previous systematic review [6], get the survey reviewed by another researchers, and followed by conducting a pilot testing (i.e., surveying 10% of the current participants). The pilot survey helped us to have more confident that we have covered all the important tasks within the development process of BBS. The respondents who participated in the pilot survey have not suggested any clarifications to be revised.

### 7.3 External Validity

External validity refers to applying generalization of the study results. Our questionnaire was developed based on our previous systematic review [6], which tended to identify commonly grounded tasks related to BBS development process reported in the literature. However, it is likely that we have missed some other tasks, especially less cited tasks, that could be important for inclusion in the framework. We do not claim that the framework is capable to manifest all necessary tasks for all BBS development scenarios. In addition, due to the fact that we purposefully selected random samples for the survey, the survey findings may be limited to those who participated in this survey and could generate biased results. Moreover, we attempted to have participants from a range of backgrounds in our limited sample size. Yet, we cannot claim the generality of our research findings. However, when considering the maximum variation sampling rule [41] that suggests varying opinions and experiences to avoid attrition bias, we ensured targeting as diverse a population as possible in regard to countries (i.e., from 36 countries distributed across 6 continents), 20 different blockchain domains, seniority levels, and years of experience. Such a diversity of participants can be considered as good representative to answer to our RQs. It's worth mentioning that we needed to identify experts who would be active in BBS development to participate in our survey. We contacted many experts and development companies but only heard that they could not share their experience with entities which are external to their companies due to privacy, regulatory issues, and intellectual property matter. We are aware that some challenges related to the tasks may not have been shared by the respondents or have been misrepresented. Thus, we cannot firmly state that the framework and findings from our study are completely representative of the opinions about BBS development process.

# 8. Conclusions

BBS usage has proven its effectiveness in a variety of domains, especially financial and healthcare, due to the great benefits it has provided. In this study, we presented a process framework that facilitate engineering BBS. Our framework includes 26 development tasks that have been categorized into a five-phase based process framework (i.e., analysis, design, implementation, deployment, and maintenance), and then obtaining quantitative and qualitative to support the framework through Web-based survey results. We believe our study can serve as a guideline, as we presented the framework tasks that have been derived through synthesizing the literature relevant to the field of SE for blockchain. Future work may consider further investigation regarding the applicability of our proposed framework with specific domain. Such an investigation would lead to identifying the relevant software engineering tasks for a particular domain (i.e., including new tasks or excluding existing tasks).

## References

[1] M. Crosby, P. Pattanayak, S. Verma, and V. Kalyanaraman, "Blockchain technology: Beyond bitcoin," Applied Innovation, vol. 2, p. 71, 2016.

[2] ""Deloitte's 2019 global Blockchain survey-Blockchain gets down to business" avaliable at https://www2.deloitte.com/content/dam/Deloitte/se/Documents/risk/DI_2019-global-blockchain-survey.pdf [Last access: March 2024]," 2019.

[3] A. Bosu, A. Iqbal, R. Shahriyar, and P. Chakraborty, "Understanding the motivations, challenges and needs of blockchain software developers: A survey," Empirical Software Engineering, vol. 24, pp. 2636-2673, 2019.

[4] H. Chidera, "What are the Top Blockchain Developer Communities available at https://www.commudle.com/blogs/top-blockchain-developer-communities," 2024.

[5] W. Chen, Z. Xu, S. Shi, Y. Zhao, and J. Zhao, "A survey of blockchain applications in different domains," in Proceedings of the 2018 International Conference on Blockchain Technology and Application, 2018, pp. 17-21.

[6] M. Fahmideh, J. Grundy, A. Ahmed, J. Shen, J. Yan, D. Mougouei, et al., "Software engineering for blockchain based software systems: Foundations, survey, and future directions," arXiv preprint arXiv:2105.01881, 2021.

[7] J. Yli-Huumo, D. Ko, S. Choi, S. Park, and K. Smolander, "Where is current research on blockchain technology?—a systematic review," PloS one, vol. 11, p. e0163477, 2016.

[8] E. Foster and B. Towle Jr, Software engineering: a methodical approach: Auerbach Publications, 2021.

[9] S. Porru, A. Pinna, M. Marchesi, and R. Tonelli, "Blockchain-oriented software engineering: challenges and new directions," in 2017 IEEE/ACM 39th International Conference on Software Engineering Companion (ICSE-C), 2017, pp. 169-171.

[10] M. Pilkington, "Blockchain technology: principles and applications," in Research handbook on digital transformations, ed: Edward Elgar Publishing, 2016, pp. 225-253.

[11] M. Jurgelaitis, R. Butkienė, E. Vaičiukynas, V. Drungilas, and L. Čeponienė, "Modelling principles for blockchain-based implementation of business or scientific processes," in CEUR workshop proceedings: IVUS 2019 international conference on information technologies: proceedings of the international conference on information technologies, Kaunas, Lithuania, April 25, 2019, 2019, pp. 43-47.

[12] M. Swan, Blockchain: Blueprint for a new economy: " O'Reilly Media, Inc.", 2015.

[13] M. Pilkington, "Blockchain technology: principles and applications," in Research handbook on digital transformations, ed: Edward Elgar Publishing, 2016.

[14] C. Majaski, "Distributed leadger available at https://www.investopedia.com/terms/d/distributed-ledgers.asp," 2021.

[15] A. A. Monrat, O. Schelén, and K. Andersson, "A survey of blockchain from the perspectives of applications, challenges, and opportunities," IEEE Access, vol. 7, pp. 117134-117151, 2019.

[16] F. Casino, T. K. Dasaklis, and C. Patsakis, "A systematic literature review of blockchain-based applications: Current status, classification and open issues," Telematics and informatics, vol. 36, pp. 55-81, 2019.

[17] S. Demi, R. Colomo-Palacios, and M. Sánchez-Gordón, "Software engineering applications enabled by blockchain technology: A systematic mapping study," Applied sciences, vol. 11, p. 2960, 2021.

[18] N. Dimitrijević, N. Zdravkovic, and V. Milicevic, "A comparative overview on Blockchain-based applications for Software Engineering," 2022.

[19] M. Risius and K. Spohrer, "A blockchain research framework," Business & Information Systems Engineering, vol. 59, pp. 385-409, 2017.

[20] Z. Zheng, S. Xie, H.-N. Dai, X. Chen, and H. Wang, "Blockchain challenges and opportunities: A survey," International journal of web and grid services, vol. 14, pp. 352-375, 2018.

[21] I. Sommerville, "Software Engineering 7th ed. Addison-Wesley, Reading, Mass," 2004.

[22] D. H. Ingalls, "The Smalltalk-76 programming system design and implementation," in Proceedings of the 5th ACM SIGACT-SIGPLAN symposium on Principles of programming languages, 1978, pp. 9-16.

[23] S. L. Pfleeger, "Albert Einstein and empirical software engineering," Computer, vol. 32, pp. 32-38, 1999.

[24] G. Cugola and C. Ghezzi, "Software Processes: a Retrospective and a Path to the Future," Software Process: Improvement and Practice, vol. 4, pp. 101-123, 1998.

[25] A. Fuggetta, "Software process: a roadmap," in Proceedings of the Conference on the Future of Software Engineering, 2000, pp. 25-34.

[26] P. Chakraborty, R. Shahriyar, A. Iqbal, and A. Bosu, "Understanding the software development practices of blockchain projects: a survey," in Proceedings of the 12th ACM/IEEE international symposium on empirical software engineering and measurement, 2018, pp. 1-10.

[27]	M. Marchesi, L. Marchesi, and R. Tonelli, "An agile software engineering method to design blockchain applications," in Proceedings of the 14th Central and Eastern European Software Engineering Conference Russia, 2018, pp. 1-8.
[28]	G. Destefanis, M. Marchesi, M. Ortu, R. Tonelli, A. Bracciali, and R. Hierons, "Smart contracts vulnerabilities: a call for blockchain software engineering?," in 2018 International Workshop on Blockchain Oriented Software Engineering (IWBOSE), 2018, pp. 19-25.
[29]	A. Pinna, G. Baralla, M. Marchesi, and R. Tonelli, "Raising sustainability awareness in agile blockchain-oriented software engineering," in 2021 IEEE International Conference on Software Analysis, Evolution and Reengineering (SANER), 2021, pp. 696-700.
[30]	M. Fahmideh, J. Grundy, A. Ahmad, J. Shen, J. Yan, D. Mougouei, et al., "Engineering Blockchain-based Software Systems: Foundations, Survey, and Future Directions," ACM Computing Surveys, vol. 55, pp. 1-44, 2022.
[31]	S. Demi, R. Colomo-Palacios, and M. Sanchez-Gordon, "Software engineering applications enabled by blockchain technology: A systematic mapping study," Applied sciences, vol. 11, p. 2960, 2021.
[32]	A. Vacca, A. Di Sorbo, C. A. Visaggio, and G. Canfora, "A systematic literature review of blockchain and smart contract development: Techniques, tools, and open challenges," Journal of Systems and Software, vol. 174, p. 110891, 2021.
[33]	C. Lal and D. Marijan, "Blockchain testing: Challenges, techniques, and research directions," arXiv preprint arXiv:2103.10074, 2021.
[34]	M. Li, J. Weng, A. Yang, W. Lu, Y. Zhang, L. Hou, et al., "CrowdBC: A blockchain-based decentralized framework for crowdsourcing," IEEE Transactions on Parallel and Distributed Systems, vol. 30, pp. 1251-1266, 2018.
[35]	U. Farooq, M. Ahmed, S. Hussain, F. Hussain, A. Naseem, and K. Aslam, "Blockchain-based software process improvement (bbspi): An approach for smes to perform process improvement," IEEE Access, vol. 9, pp. 10426-10442, 2021.
[36]	D. Ulybyshev, M. Villarreal-Vasquez, B. Bhargava, G. Mani, S. Seaberg, P. Conoval, et al., "(WIP) blockhub: Blockchain-based software development system for untrusted environments," in 2018 IEEE 11th International Conference on Cloud Computing (CLOUD), 2018, pp. 582-585.
[37]	R. J. C. Bose, K. K. Phokela, V. Kaulgud, and S. Podder, "Blinker: A blockchain-enabled framework for software provenance," in 2019 26th Asia-Pacific Software Engineering Conference (APSEC), 2019, pp. 1-8.
[38]	M. J. H. Faruk, S. Subramanian, H. Shahriar, M. Valero, X. Li, and M. Tasnim, "Software engineering process and methodology in blockchain-oriented software development: A systematic study," in 2022 IEEE/ACIS 20th International Conference on Software Engineering Research, Management and Applications (SERA), 2022, pp. 120-127.
[39]	A. Pinsonneault and K. Kraemer, "Survey research methodology in management information systems: an assessment," Journal of management information systems, vol. 10, pp. 75-105, 1993.
[40]	"An online web-based survey for exploration of Blockchain-based system development study available at https://forms.gle/VMf4ZPD2JwfoJdGX7," 2022.
[41]	M. Q. Patton, Qualitative evaluation and research methods: SAGE Publications, inc, 1990.
[42]	M. Rosenblatt, "A central limit theorem and a strong mixing condition," Proceedings of the national Academy of Sciences, vol. 42, pp. 43-47, 1956.
[43]	F. F. E. E. Axel Buchner, Albert-Georg Lang, "G*Power: Statistical Power Analyses for Windows and Mac avaliable at http://www.psychologie.hhu.de/arbeitsgruppen/allgemeine-psychologie-und-arbeitspsychologie/gpower.html," 2007.
[44]	J. Cohen, "Statistical power analysis," Current directions in psychological science, vol. 1, pp. 98-101, 1992.
[45]	D. S. Cruzes and T. Dyba, "Recommended steps for thematic synthesis in software engineering," in Empirical Software Engineering and Measurement (ESEM), 2011 International Symposium on, 2011, pp. 275-284.
[46]	V. Braun and V. Clarke, "Using thematic analysis in psychology," Qualitative research in psychology, vol. 3, pp. 77-101, 2006.
[47]	M. B. Miles and A. M. Huberman, Qualitative data analysis: An expanded sourcebook: sage, 1994.
[48]	I. Sommerville and J. Ransom, "An empirical study of industrial requirements engineering process assessment and improvement," ACM Transactions on Software Engineering and Methodology (TOSEM), vol. 14, pp. 85-117, 2005.
[49]	M. H. Miraz and M. Ali, "Blockchain enabled smart contract based applications: Deficiencies with the software development life cycle models," arXiv preprint arXiv:2001.10589, 2020.